\documentclass[amsmath,amssymb,10pt,prl,showkeys]{revtex4}

\usepackage{graphicx}
\usepackage{bm}

\usepackage{color}

\begin{document}
\title{Measurement-induced strong Kerr nonlinearity for weak quantum states of light}

\author{Luca S. Costanzo$^{1,2}$, Antonio S. Coelho$^{3}$, Nicola Biagi$^{1,2}$, Jarom\'{i}r Fiur\'{a}\v{s}ek$^{4}$, Marco Bellini$^{1,2}$, and Alessandro Zavatta$^{1,2^\dagger}$}

\affiliation{$^{1}$Istituto Nazionale di Ottica (INO-CNR), L.go E. Fermi 6, 50125 Florence, Italy\\
$^{2}$LENS and Department of Physics, University of Firenze, 50019 Sesto Fiorentino, Florence, Italy\\
$^{3}$Departamento de Engenharia Mec\^anica, Universidade Federal do Piau\'i, 64049-550, Teresina, PI, Brazil\\
$^{4}$Department of Optics, Palack\'{y} University, 17. listopadu 1192/12, CZ-771 46 Olomouc, Czech Republic}

\bigskip
\date{\today}

\begin{abstract}
Strong nonlinearity at the single photon level represents a crucial enabling tool for optical quantum technologies. Here we report on experimental implementation of a strong Kerr nonlinearity by measurement-induced quantum operations on weak quantum states of light. Our scheme coherently combines two sequences of single photon addition and subtraction to induce a nonlinear phase shift at the single photon level. We probe the induced nonlinearity with weak coherent states and characterize the output non-Gaussian states with quantum state tomography. The strong nonlinearity is clearly witnessed as a change of sign of specific off-diagonal density matrix elements in Fock basis.
\end{abstract}

\maketitle

Nonlinear optical interactions represent a major tool for generation and manipulation of optical fields in both classical and quantum domains, and they form a basis of countless photonics devices. The envisioned applications of nonlinear interactions in quantum optics and optical quantum information processing often require strong nonlinear coupling between single photons. However, such ultrastrong nonlinear interactions are not readily available, because the typical nonlinearities of common non-resonant optical media are many orders of magnitude weaker than what is required to achieve an appreciable nonlinearity at the single-photon level. 
A significant progress towards giant optical nonlinearities at the few-photon level has been made during recent years due to intensive experimental and theoretical efforts \cite{Fushman08,Peyronel12,Venkataraman13,Chang14,Volz14,Feizpour15,Ralph15,Hacker16,Gorniaczyk16,Snijders16,Bennett16,Tiarks16,Tresp16,Das16,Beck16}. However, those experimental approaches are extremely complex and challenging as they require specially tailored media with enhanced nonlinearities such as clouds of ultracold atoms. Moreover, several works pointed out that the very nature of light-matter interaction may prevent achievement of a sufficiently strong Kerr nonlinearity in certain configurations for weak quantum optical fields \cite{Shapiro06,Gea-Banacloche10}.

In 2001, Knill, Laflamme, and Milburn in their landmark paper showed that effective nonlinear interactions at the single-photon level can be implemented with the use of optical interference, single photon detection and auxiliary single photons \cite{KLM01}. In this approach, the single photon detection provides the desired nonlinearity. The resulting linear optical quantum gates  are generally probabilistic, as implied by the fact that they are driven by quantum measurements, but their success probability can be boosted arbitrarily close to $1$ by using more ancilla photons and more complex interferometric schemes \cite{KLM01,Franson02}. This concept has triggered an immense amount of theoretical and experimental work, which lead to demonstration of various quantum gates for single-photon qubits \cite{Kok07,Ralph10}.
This concept has also been extended to Gaussian operations on continuous-variable states of light \cite{Filip05}, where it was demonstrated that a squeezing operation can be implemented using an auxiliary source of squeezed states, interference, homodyne detection, and feedforward \cite{Yoshikawa07}. Similarly, a quantum-noise limited phase insensitive amplification has been implemented solely by a homodyne detection and feedforward \cite{Josse06}.
 
A fundamental nonlinear interaction is represented by a Kerr nonlinearity, which leads to dependence of the refractive index on the intensity of light that propagates through the nonlinear medium. At the quantum level, this nonlinearity is described by a Hamiltonian which is a quadratic function of the photon number operator $\hat{n}$,
\begin{equation}
\hat{H}=\hbar \kappa \hat{a}^{\dagger 2} \hat{a}^2 =\hbar \kappa\hat{n}(\hat{n}-1).
\end{equation}
The resulting unitary transformation of the quantum state of the optical mode is diagonal in Fock basis, which means that each Fock state $|n\rangle$ acquires a phase shift which is a non-linear function of $n$,
\begin{equation}
|n\rangle \rightarrow e^{i\Phi \hat{n}(\hat{n}-1)}|n\rangle,
\end{equation}
where $\Phi=\kappa t$. Strong Kerr nonlinearity with $\Phi \approx 1$ would enable e.g. generation of macroscopic superpositions of coherent states \cite{Yurke86}, implementation of entangling quantum gates for universal quantum computing \cite{Milburn89}, and complete Bell state measurement in quantum teleportation \cite{Vitali00,Paris00}. 

Here we report on the experimental implementation of a strong Kerr nonlinearity by measurement-induced quantum operations on weak quantum states of light. Specifically, we emulate this interaction on the smallest non-trivial subspace spanned by the vacuum, single-photon and two-photon states, $|0\rangle$, $|1\rangle$ and $|2\rangle$. In this subspace, the Kerr interaction transforms a generic input state according to 
\begin{equation}
e^{-i\hat{H}t/\hbar}(c_0|0\rangle+c_1|1\rangle+c_2|2\rangle)=c_0|0\rangle+c_1|1\rangle+e^{-2i\Phi}c_2|2\rangle.
\end{equation} 

\begin{figure}[!t!]
\includegraphics[width=0.95\linewidth,angle=0]{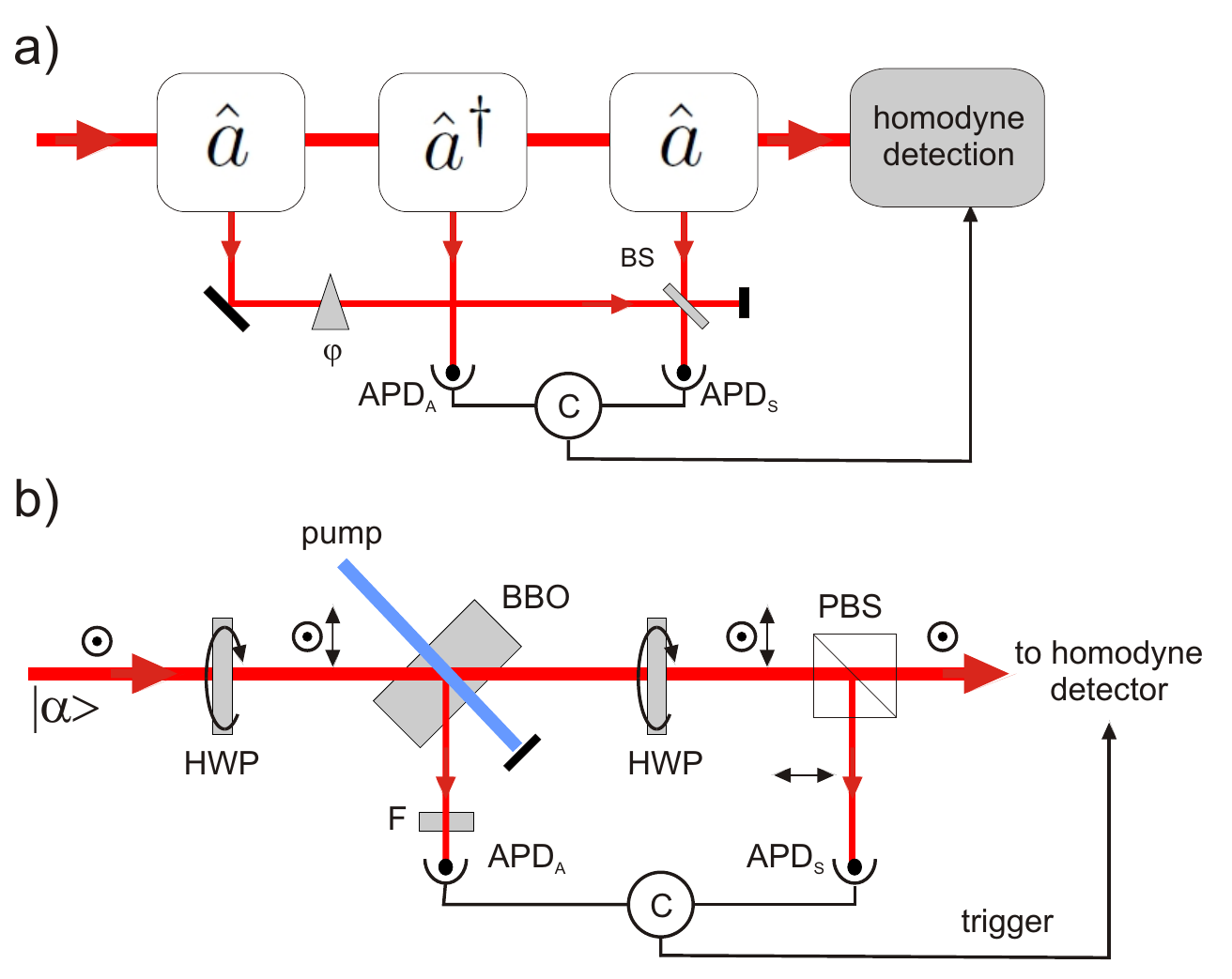}
\caption{Experimental setup. a) Schematic view of the main blocks used for implementing arbitrary coherent superpositions of sequences of conditional photon additions and subtractions. b) Detailed view of the experimental setup, based on different polarization modes for interferometric stability. All symbols are defined in the text.} \label{fig:setup}
\end{figure}

We target a Kerr nonlinearity with $\Phi=\pi/2$, which induces a $\pi$-phase shift of the two-photon Fock state with respect to states $|0\rangle$ and $|1\rangle$. Up to a linear $\pi$-phase shift which flips the sign of odd Fock states, and an unimportant overall phase factor $-1$, this is equivalent to a $\pi$-phase shift in the amplitude of the vacuum state on the three-dimensional subspace considered, i.e.,
\begin{equation}
c_0|0\rangle+c_1|1\rangle+c_2|2\rangle\rightarrow -c_0|0\rangle+c_1|1\rangle+c_2|2\rangle.
\label{Kerr}
\end{equation} 
The change of sign in the amplitude of the vacuum component is thus the signature of the strong Kerr nonlinearity that we wish to demonstrate in our experiment. We do this by using weak coherent states as the input of an approximate Kerr Hamiltonian implemented with coherent superpositions of sequences of photon additions and subtractions \cite{ZavattaPO10} (see Fig.~\ref{fig:setup}(a) for a schematic view of the experiment).

Conditional photon subtraction is usually implemented with the help of a highly unbalanced beam splitter and a single-photon detector, whose click heralds successful photon subtraction, which is mathematically described by the annihilation operator $\hat{a}$ \cite{Ourjoumtsev06,Neergaard-Nielsen06,Parigi07}. To conditionally add a single photon, the input light beam is sent to a signal input port of a non-linear crystal, where parametric down-conversion takes place. Detection of a photon in the output idler port of the crystal heralds the generation of a twin photon in the signal mode, and this operation can be described by the action of a creation operator $\hat{a}^\dagger$ \cite{Zavatta04}. 
Since the creation and annihilation operators do not commute, the sequences $\hat{a}\hat{a}^\dagger$ and $\hat{a}^\dagger \hat{a}$ represent two different operations, as verified experimentally \cite{Parigi07}. 

An interferometric scheme allows us to implement arbitrary coherent superpositions of these two elementary sequences, $A \hat{a} \hat{a}^{\dagger}+B \hat{a}^\dagger \hat{a}$ by erasing the information whether the photon subtraction took place before or after the photon addition. Taking into account that $\hat{a}^\dagger \hat{a}=\hat{n}$ and $\hat{a} \hat{a}^\dagger=\hat{n}+1$, we can thus design an arbitrary operation which is a linear function of the photon number operator, $V(\hat{n})=(A+B)\hat{n}+A$.
We already demonstrated the reliability of such a scheme for the first direct experimental verification of the bosonic commutation rules \cite{Zavatta09} and for the realization of a quantum state orthogonalizer \cite{Coelho16}.

Here, we want to conditionally implement the gate of Eq.~(\ref{Kerr}) by means of the transformation $V(n)$, therefore we need to set $V(1)/V(0)=-1$ and $V(2)/V(1)=1$. Although these conditions cannot be satisfied for any $A$ and $B$, we can make our task feasible by allowing for a simultaneous noiseless amplification \cite{Zavatta11} of the output state,
\begin{equation}
c_0|0\rangle+c_1|1\rangle+c_2|2\rangle\rightarrow -c_0|0\rangle+gc_1|1\rangle+g^2c_2|2\rangle,
\label{gKerr}
\end{equation}
where $g>1$ is a gain factor.  
The equations $V(1)/V(0)=-g$ and $V(2)/V(1)=g$ now possess a non-trivial solution $B/A=-3-\sqrt{2}$,  which yields $g=\sqrt{2}+1$. 
If wishing to achieve the nonlinearity of Eq.~(\ref{Kerr}) without amplification, one could either use two photon subtractions and additions instead of one \cite{Fiurasek09}, or the output state of Eq.~(\ref{gKerr}) could be noiselessly attenuated \cite{Micuda12,Gagatsos14} with the help of a beam splitter with amplitude transmittance $t=1/g$, a highly efficient single-photon detector, and conditioning on observation of no photons at the auxiliary output port of the beam splitter. Note, however, that the additional noiseless amplification in Eq.~(\ref{gKerr})  does not spoil the signatures of nonlinearity. On the contrary, it is actually beneficial, because we intend to probe the quantum operation with weak coherent states and the amplification makes the observed nonlinear effect even more visible.

The experiment requires the precise and stable setting of the relative weights and phases of the operator superposition to implement the desired conditional transformation, which is then tested by using a set of weak coherent states as the input. The resulting output states are subjected to balanced homodyne detection and finally analyzed via a full quantum state tomographic reconstruction.

\begin{figure*}[!t!]
   \includegraphics[width=.9\linewidth,angle=0]{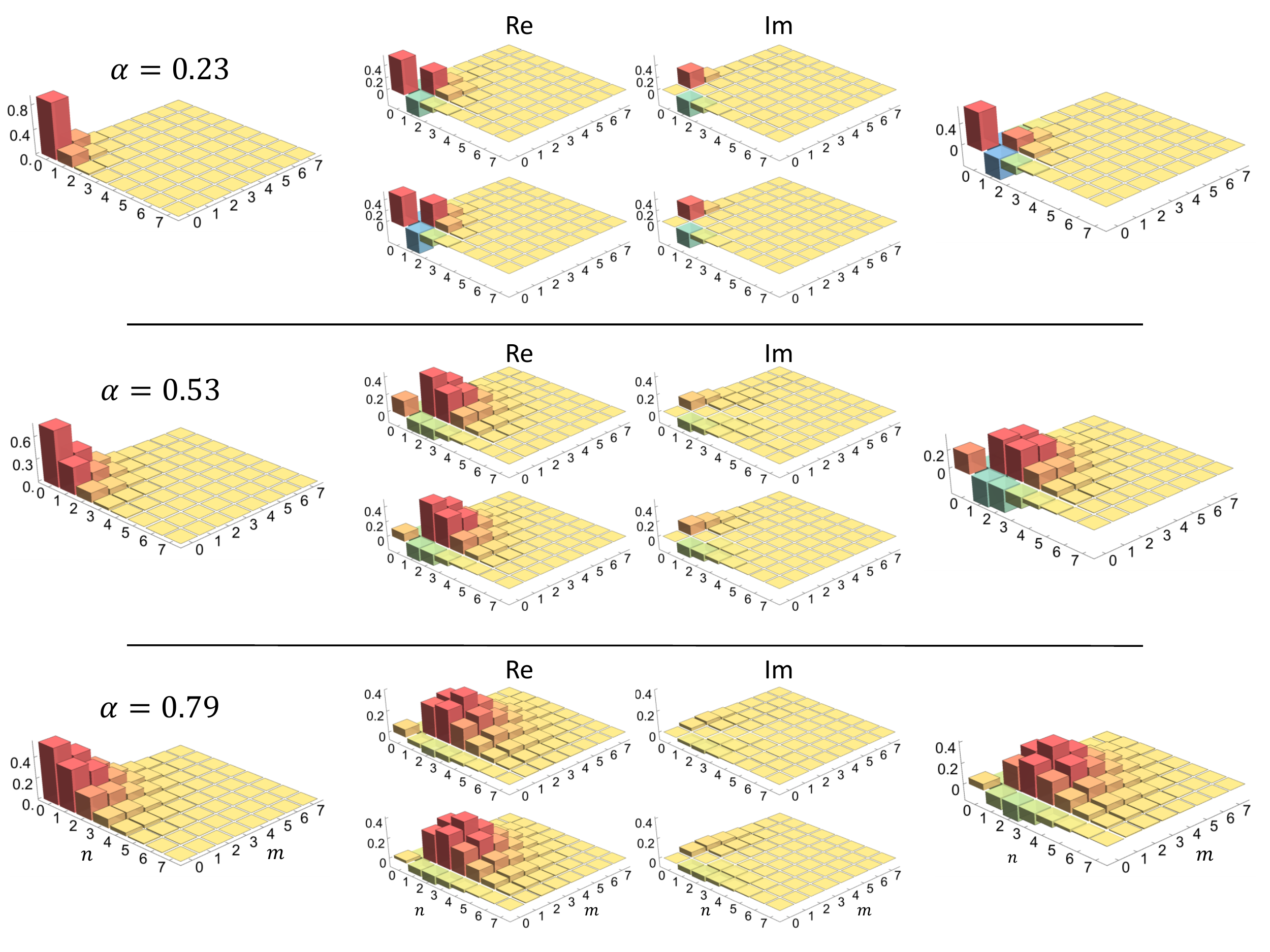}
\caption{Reconstructed density matrices of input coherent states and output states after the emulated Kerr nonlinear interaction. The left column shows the real part of the reconstructed density matrices of the input coherent states (the imaginary part is negligible here). For each input state, the two central columns show the real and imaginary parts for the reconstructed output states (upper plots), together with those calculated from a best fit of the parameters in the applied $V(\hat{n})$ transformation (lower plots). The corresponding fidelities are  $F=0.88, 0.86, 089$ for $\alpha=0.23, 0.53, 0.79$, respectively. Finally, the right column shows the expected output states (containing no imaginary parts) that one would obtain from the ideal $V(\hat{n})$ transformation with $B/A=-3-\sqrt{2}$.} \label{fig:densmat}
\end{figure*}

Differently from the previous realizations based on a bulk fiber optic interferometer \cite{Zavatta09}, here we choose to exploit the polarization degree of freedom for realizing a more compact and phase-stable setup that enables us to meet the stringent requirements on the operator superposition.
Figure \ref{fig:setup}(b) shows a more detailed view of the experimental setup. A mode-locked Ti:sapphire laser emitting 1.5 ps long pulses at 786 nm is frequency-doubled to pump the 3-mm long bulk $\beta$-barium borate (BBO) crystal for frequency-degenerate, non-collinear parametric down-conversion (PDC). A small part of the laser emission constitutes, after proper attenuation, the input coherent states $|\alpha\rangle$ that are injected in the signal mode of the PDC crystal for conditional photon addition, heralded by clicks in the avalanche photodiode (APD$_\text{A}$) placed after narrow spectral and spatial filters (F) in the idler mode. Two half-wave plates (HWP) are placed in the signal mode, respectively before and after the PDC crystal. If the two plates are rotated by very small angles, then each HWP transfers a tiny portion of the signal light into the orthogonal polarization component, which does not contribute to PDC and is detected after a polarizing beam splitter (PBS) by another avalanche photodiode (APD$_\text{S}$). A click of this detector thus heralds subtraction of a single photon from the signal pulse that could have occurred either before or after photon addition. The relative (real) amplitudes of the terms $\hat{a} \hat{a}^\dagger$ and $\hat{a}^\dagger \hat{a}$ in the superposition heralded by a coincidence between APD$_\text{A}$ and APD$_\text{S}$ can be finely controlled by rotating the two HWPs by different angles.

In order to prepare complex superpositions, one might add a quarter wave plate right after the first HWP, but this is not necessary for the present experiment. However, we do add a compensating BBO crystal (not shown in the figure for simplicity) to cancel the spatiotemporal walk-off of the two orthogonally polarized pulses (and thus the partial distinguishability between the two 'subtraction' beams) due to the birefringence of the PDC crystal.

This polarization-based scheme is inherently interferometrically stable because the main polarization component of the signal light pulse and the orthogonally-polarized subtraction pulses follow the same path until they are separated by the PBS. After the PBS, the signal pulse is directed towards a time-domain balanced homodyne detector \cite{Zavatta02,Zavatta04b} using another portion of the laser emission as the reference local oscillator pulses. The detector thus acquires phase-dependent quadrature data corresponding to the input coherent states $|\alpha\rangle$ in its free-running mode, whereas APD$_\text{A}$-APD$_\text{S}$ coincident clicks herald quadrature measurements on the conditional output states. 

A full tomographic reconstruction is performed on the input and output states for three different values of the input coherent state amplitude, $\alpha$= 0.23, 0.53 and 0.79. We use an iterative maximum likelihood procedure \cite{Hradil04,Lvovsky04,Hradil07} incorporating the effect of a finite ($\eta_{\mathrm{det}}=0.66$) detector efficiency to reconstruct the density matrices in a 8x8 space in the Fock basis. 
The reconstructed density matrices are shown in Fig.~\ref{fig:densmat} together with those calculated by applying the $V(\hat{n})$ operator on the input coherent states.

\begin{figure}[!t!]
\includegraphics[width=0.8\linewidth]{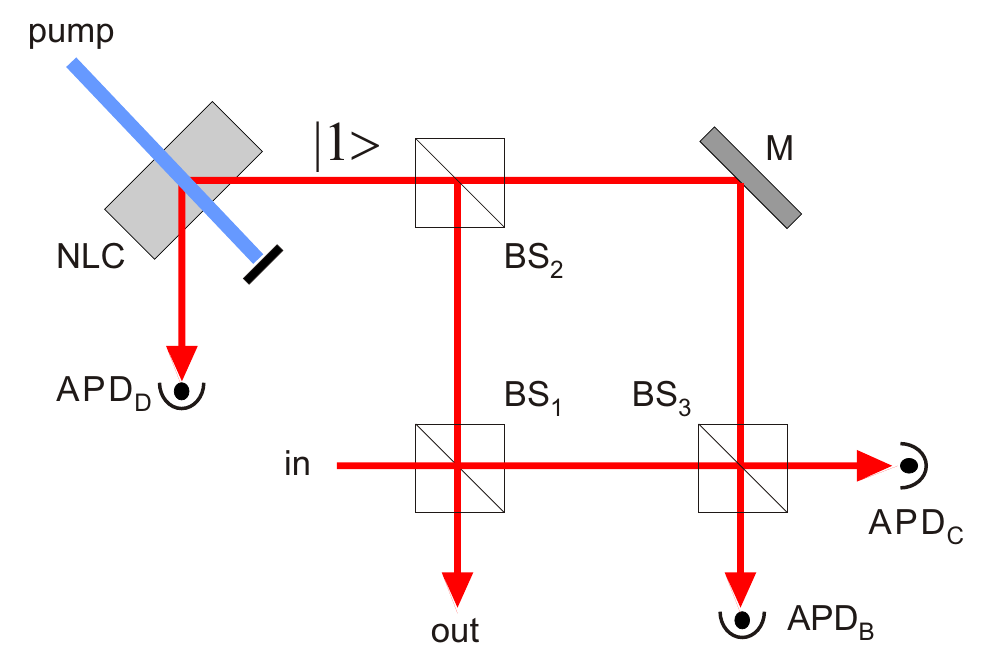}
\caption{Nonlinear sign gate proposed by Knill, Laflamme and Milburn \cite{KLM01}. The scheme consists of beam splitters (BS), single-photon detectors (APD), a mirror (M), and a nonlinear crystal (NLC) pumped by a strong laser beam (pump). The input and output signal ports are labeled by \emph{in} and \emph{out}, respectively. Successful gate operation is heralded by detection of a single photon by detectors APD$_{\mathrm{B}}$ and APD$_{\mathrm{D}}$ while no photon should be detected by APD$_{\mathrm{C}}$.}
\label{fig:KLMgate}
\end{figure}

The desired Kerr nonlinearity signature is evident in all the experimental data. All the off-diagonal terms containing a vacuum contribution are clearly negative, witnessing the expected sign change in the amplitude of the vacuum component. However, when comparing the experimental density matrices to those expected according to the $V(\hat{n})$ transformation with ideal parameters $B/A=-3-\sqrt{2}$ (rightmost column in Fig.~\ref{fig:densmat}), some discrepancy is apparent. The most notable is the appearance of a small imaginary component. We find that all the experimental results can be reproduced very well by using a single set of modified parameters in the $V(\hat{n})$ transformation, corresponding to a $B/A$ ratio of $-5.97$ and to an additional phase of about $-\pi/7$ between the two terms in the operator superposition. Such small deviations from the ideal configuration, which only marginally affect the signatures of the sought nonlinearity, are fully compatible with the delicate alignment and setting of the proper small rotation angles in the waveplates responsible for the operator superposition.

It is instructive to compare our scheme with the nonlinear sign gate proposed by Knill, Laflamme and Milburn \cite{KLM01}, which is schematically depicted in Fig.~\ref{fig:KLMgate}. The nonlinear sign gate conditionally implements the transformation (\ref{Kerr}) and requires one auxiliary input single photon. To make a fair comparison with our setup, we include in Fig.~\ref{fig:KLMgate} a  source of single photon states based on generation of correlated photon pairs in the process of spontaneous parametric down-conversion in a nonlinear crystal followed by heralding detection of the idler photon. 
The nonlinear sign gate involves interference of the input signal mode with the auxiliary single photon in a three-port interferometer composed of three unbalanced beam splitters with suitably chosen transmittances. Single photon detectors monitor the two auxiliary output ports of the interferometer and successful implementation of the gate (\ref{Kerr}) is heralded if a single photon is detected by APD$_B$ and no photon is detected by APD$_C$.

We can see that the the setup shown in Fig.~\ref{fig:KLMgate} and our scheme depicted in Fig.~\ref{fig:setup} require comparable resources, namely production of one auxiliary photon pair in a nonlinear crystal and detection of two single photons. However, the nonlinear sign gate in Fig. 3 requires perfect photon number resolving detectors for reliable operation, while with our present approach a high-fidelity implementation of the transformation (\ref{gKerr}) can be achieved even with ordinary imperfect on-off avalanche photodiodes. Note that the scheme in Fig.~\ref{fig:KLMgate} also requires projection of one output auxiliary mode onto vacuum. In our scheme in Fig.~\ref{fig:setup} this additional ingredient of projection onto vacuum could be used for noiseless attenuation of the output signal mode, which would allow us to implement the exact nonlinear sign gate (\ref{Kerr}) instead of its amplified version (\ref{gKerr}). 



In conclusion, we have successfully demonstrated a heralded scheme based on sequences and superpositions of single-photon additions and subtractions that is able to emulate the action of a strong Kerr nonlinearity on quantum states of light. We tested it on various weak coherent states and showed the clear expected signature of the operation, consisting in the selective change in the sign of the vacuum component. 
In contrast to linear-optical quantum gates for single-photon qubits \cite{Kok07,Ralph10}, our scheme is not restricted to single-photon inputs, but we have shown it to work for arbitrary superpositions of Fock states while preserving the quantum coherence. Moreover, the scheme is not based on post-selection, but the successful implementation of the operation (\ref{gKerr}) is heralded solely by measurements on auxiliary modes, thus making the output state available for further processing and applications.

L.S.C., N. B., M. B. and A. Z. gratefully acknowledge the support of Ente Cassa di Risparmio di Firenze and of the Italian Ministry of Education, University and Research (MIUR), under the 'Progetto Premiale: QSecGroundSpace'. J.F. acknowledges financial support by the Czech Science Foundation (GB14-36681G).

\end{document}